\documentclass[fleqn,twoside]{article}
\usepackage{espcrc2}

\usepackage{graphicx}

\title{
  Chiral extrapolation of light-light and heavy-light decay constants
  in unquenched QCD\thanks{
    Talk presented by S.~Hashimoto}
  }

\author{
  JLQCD Collaboration:
  S.~Hashimoto
  \address{
    High Energy Accelerator Research Organization (KEK), 
    Tsukuba 305-0801, Japan
    },
  S.~Aoki
  \address{
    Institute of Physics, University of Tsukuba, 
    Tsukuba 305-8571, Japan.
    },
  M.~Fukugita
  \address{
    Institute for Cosmic Ray Research,
    University of Tokyo, Kashiwa, Chiba 277-8582, Japan.
    }, 
  K-I.~Ishikawa$^{\rm b,}$\address{
    Center for Computational Physics, 
    University of Tsukuba, Tsukuba 305-8577, Japan.
    }, 
  N.~Ishizuka$^{\rm b}$
  Y.~Iwasaki$^{\rm b,d}$,
  K.~Kanaya$^{\rm b,d}$, 
  T.~Kaneko$^{\rm a}$,
  Y.~Kuramashi$^{\rm a}$, 
  M.~Okawa
  \address{
    Department of Physics,
    Hiroshima University, Higashi-Hiroshima 739-8526, Japan.
    },
  N.~Tsutsui$^{\rm a}$,
  A.~Ukawa$^{\rm b,d}$,
  N.~Yamada$^{\rm a}$,
  and 
  T.~Yoshi\'e$^{\rm b,d}$.
} 

\begin{document}

\begin{abstract}
  We test the one-loop chiral perturbation theory formula
  on unquenched lattice data of pseudoscalar meson decay
  constants.
  The chiral extrapolation including the effect of the
  chiral logarithm is attempted and its uncertainty is
  discussed. 
\end{abstract}

\maketitle

\section{Introduction}
In the unquenched QCD simulations using the Hybrid Monte
Carlo algorithm, the computational cost rapidly grows as the
chiral limit of sea quark is approached.
In practical simulations with Wilson-type fermions the sea
quark mass is limited to be heavier than $\sim m_s/2$.
To obtain the physical results for $u$ and $d$ quarks,
therefore, the chiral extrapolation is indispensable.

In the chiral extrapolation the chiral perturbation theory
(ChPT) may be used to decide the functional form, as it is
an effective theory valid for low energy QCD.
Once the available lattice data are confirmed to be
consistent with ChPT, the extrapolation to the $u$ and $d$
quark masses using the ChPT formula is justified. 
The practical question is, then, whether the lattice results
could reproduce the sea quark mass dependence, especially
the chiral logarithm, predicted by ChPT.

In this talk, we present the sea quark mass dependence of
the pseudoscalar meson decay constant obtained in unquenched
QCD, and compare them with the one-loop ChPT prediction.
The unquenched simulations are done using the standard gauge
and nonperturbatively $O(a)$-improved fermion action 
at $\beta$ = 5.2 
with sea quark masses corresponding to the pion mass
in the range 550--1000~MeV.
Further details of the simulation are discussed in a
separate talk \cite{Kaneko_lat2002}.

We also discuss the uncertainty associated with the chiral
extrapolation taking light-light and heavy-light decay
constants as examples.

\section{Test of the ChPT formula}

In full QCD the ChPT predicts a specific functional
dependence of physical quantities on the quark mass,
\textit{i.e.} the chiral logarithm, at the one-loop order. 
For $N_f$ flavors of degenerate quarks with a mass $m_S$, 
the pseudoscalar meson decay constant $f_{\mathrm{SS}}$ is
given by \cite{Gasser_Leutwyler} 
\begin{equation}
  \label{eq:ChPT_PSdecayconst}
    \frac{f_{SS}}{f}
    = 1 - \frac{N_f}{2} y_{\mathrm{SS}}\ln y_{\mathrm{SS}}
    + \frac{y_{\mathrm{SS}}}{2} [
    \alpha_5 + N_f \alpha_4
    ],
\end{equation}
with $y_{\mathrm{SS}}=2B_0m_S/(4\pi f)^2$.  
While the low energy constants $\alpha_i$ are unknown
parameters, the chiral log term $y_{\mathrm{SS}}\ln
y_{\mathrm{SS}}$ appears with a definite coefficient
depending only on the number of flavors. 


\begin{figure}[t]
  \begin{center}
    \leavevmode
    \includegraphics*[width=6.5cm,clip]{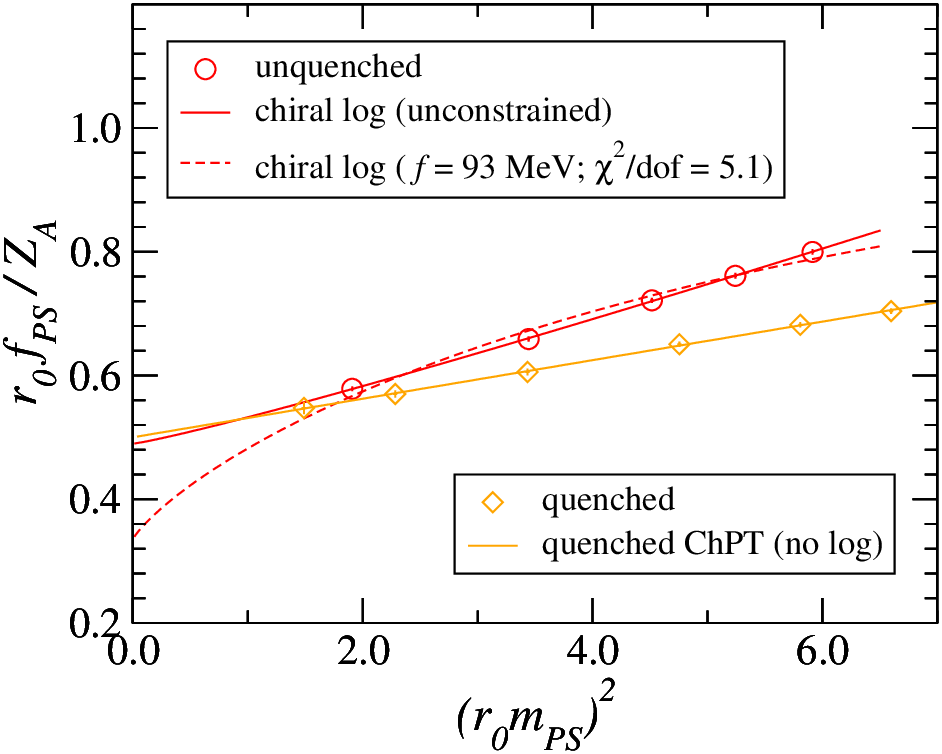}
  \end{center}
  \vspace*{-12mm}
  \caption{
    Fit of pseudoscalar meson decay constant with the one-loop
    ChPT formula. 
    }
  \vspace*{-4mm}
  \label{fig:fpi_vs_mpi2}
\end{figure}

Figure~\ref{fig:fpi_vs_mpi2} shows the lattice results 
together with fitting curves.
If we leave the coefficient of the chiral log term as a free
parameter, the fit result is consistent with zero (solid
line), while the fit with the fixed coefficient $N_f/2$
gives a bad $\chi^2/\mathrm{dof}$ (dashed line).
A similar observation is obtained for the PCAC relation
$M_{SS}^2/2m_S$ \cite{Aoki:2001yr}.

\begin{figure}[t]
  \begin{center}
    \leavevmode
    \includegraphics*[width=6.9cm,clip]{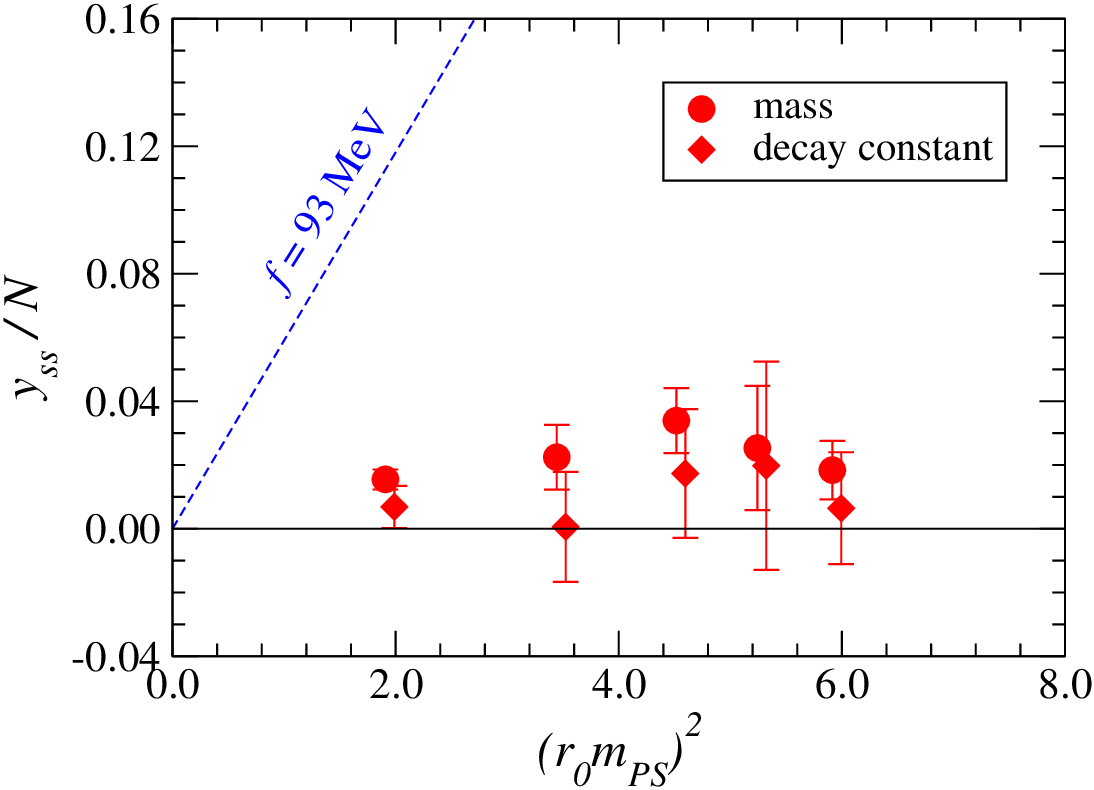}
  \end{center}
  \vspace*{-12mm}
  \caption{
    Ratio test of the partially quenched ChPT formula.
    }
  \vspace*{-4mm}
  \label{fig:ratio_test}
\end{figure}

Partially quenched ChPT \cite{Sharpe:1997by,Golterman:1998st}
may be used to explicitly test the presence of the chiral
logarithm. 
For non-degenerate pions composed of quarks with mass $m_S$
and $m_V$ ($V$ denotes a valence quark), the low energy
constants cancel out in the double ratios
\begin{eqnarray}
  \label{eq:ratio_test_PCAC}
  \frac{
    \left(\frac{M_{VS}^2}{m_V+m_S}\right)^2
  }{
    \frac{M_{SS}^2}{2m_S}\cdot\frac{M_{VV}^2}{2m_V}
  } =
  1 + \frac{y_{SS}}{N_f} t,
  \\
  \label{eq:ratio_test_decay_constant}
  \frac{f_{VS}}{\sqrt{f_{SS}f_{VV}}} =
  1 - \frac{y_{SS}}{4N_f} t,
\end{eqnarray}
with 
$t \equiv \ln (y_{VV}/y_{SS}) + 1 - (y_{VV}/y_{SS})$,
and only the chiral log terms remain.
A parameter $y_{SS}/N_f$ obtained as a coefficient of $t$ in 
(\ref{eq:ratio_test_PCAC}) and in
(\ref{eq:ratio_test_decay_constant}) are plotted as a
function of $M_{SS}^2$ in Figure~\ref{fig:ratio_test}.
We find that the results are much smaller than the
prediction of the partially quenched ChPT shown by a steep
dashed line. 

\section{Uncertainty in the chiral extrapolation}
Since the lattice data do not support the presence of the
chiral logarithm for the sea quark masses used in the
simulation, 
the chiral extrapolation using the one-loop ChPT formula is
not fully justified.
Instead, we consider several possible functional forms to
approach the chiral limit and discuss their associated
uncertainty. 

\begin{figure}[t]
  \begin{center}
    \leavevmode
    \includegraphics*[width=6.5cm,clip]{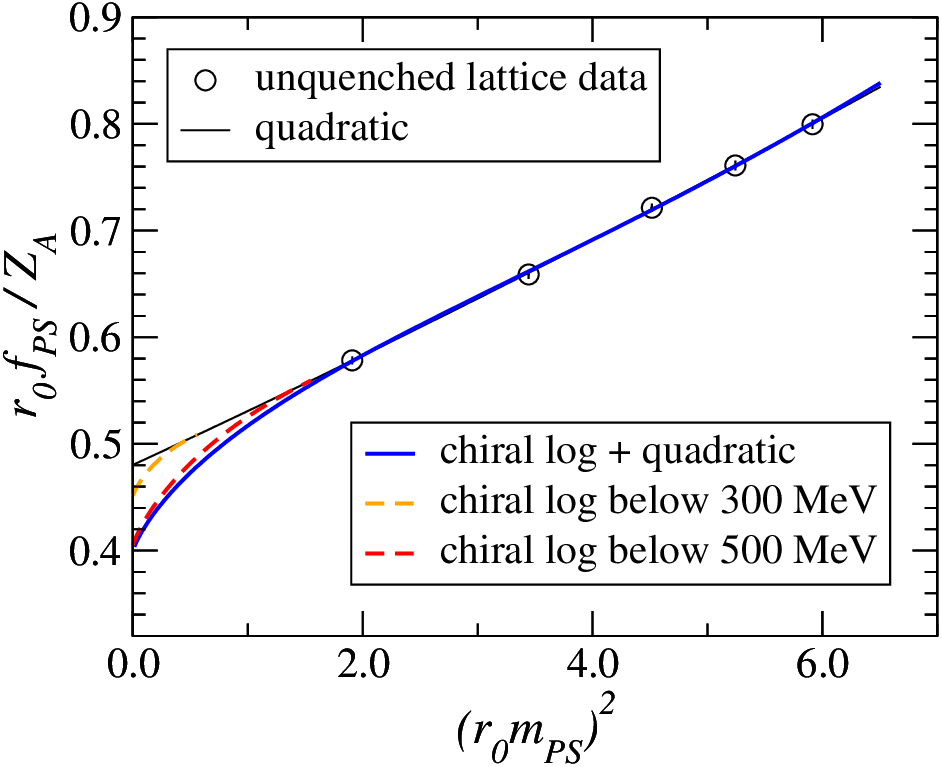}
  \end{center}
  \vspace*{-12mm}
  \caption{
    Uncertainty in the chiral extrapolation.
    }
  \vspace*{-4mm}
  \label{fig:fpi_chiral_log_below_m}
\end{figure}

Our observation suggests that the mass region
where the chiral logarithm becomes important is around or
below 500~MeV, and the ChPT ceases to converge above that
scale.  
Then, a possible way to extrapolate the data including the
effect of the chiral logarithm is to use a polynomial fit
(quadratic fit, for example) above some energy scale $\mu$,
and then switch to the one-loop ChPT formula below $\mu$.
An example is shown in
Figure~\ref{fig:fpi_chiral_log_below_m} for
$\mu$ = 300 and 500~MeV.
The limit of $\mu$ = 0~MeV corresponds to the usual
polynomial fit.
Since the scale $\mu$ is unknown, the variation of several
fit curves, about $\pm$~10\% in the chiral limit, 
should be taken as systematic uncertainty.

Another possible functional form suggested by the
Adelaide-MIT group \cite{Detmold:2001jb} is the one-loop
ChPT with a hard momentum cutoff $\mu$.
It amounts to replace the chiral log term 
$m_\pi^2\ln(m_\pi^2/\mu^2)$
by 
$m_\pi^2\ln(m_\pi^2/(m_\pi^2+\mu^2))$.
Changing the unknown ``cutoff'' scale $\mu$ from 0 to 1~GeV, 
we obtain the similar size of uncertainty in the chiral
limit.

The SU(3) breaking ratio in the decay constant $f_K/f_\pi$
may be obtained with a fit to partially quenched lattice
data. 
Our result from the above fit functions changes from
1.167(3) (quadratic fit, corresponding to $\mu$ = 0~MeV) or
1.190(3) (Adelaide-MIT fit, $\mu$ = 500~MeV) to 
1.276(7) (chiral log plus quadratic, corresponding to
$\mu=\infty$~MeV).  
Although it is clear that our two-flavor QCD result is
significantly higher than the quenched result
1.081(5)(17) \cite{Heitger:2000ay} and closer to the
physical value 1.22, the uncertainty is still sizable.

\begin{figure}[t]
  \begin{center}
    \leavevmode
    \includegraphics*[width=6.5cm,clip]{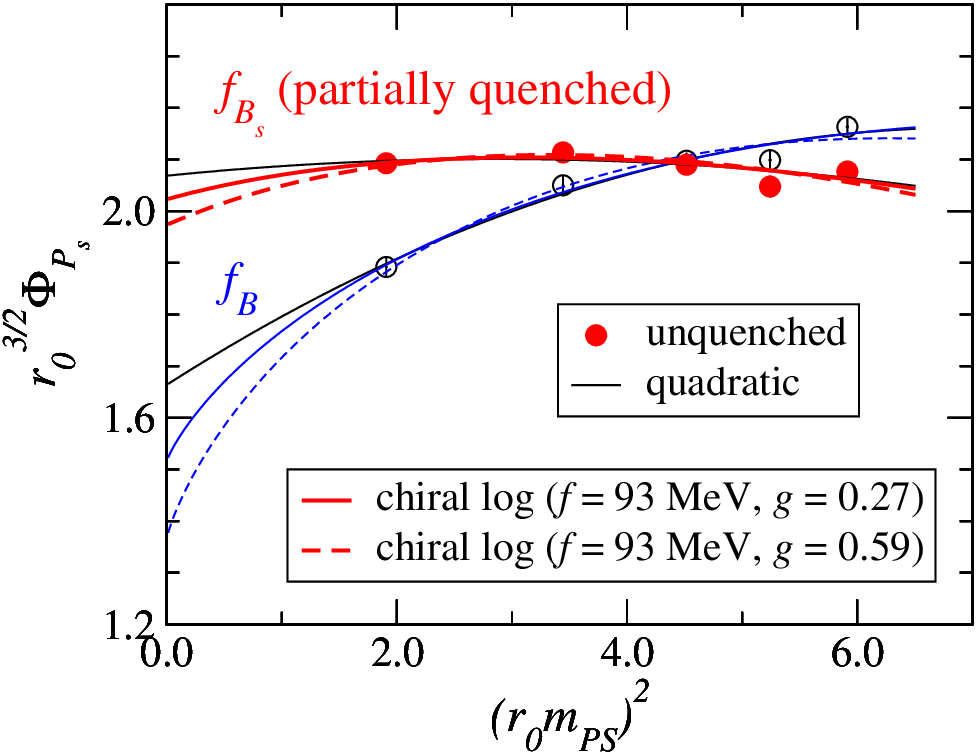}
  \end{center}
  \vspace*{-12mm}
  \caption{
    Chiral extrapolation of heavy-light decay constants.
    }
  \vspace*{-4mm}
  \label{fig:fsqrtm_strange}
\end{figure}

For the heavy-light decay constant the prediction of ChPT is
available in the heavy quark limit for
quenched, partially quenched and full QCD
\cite{Booth:1995hx,Sharpe:1996qp}.
The chiral logarithm appears with a definite coefficient but
including an additional coupling constant $g$ describing the
$B^*B\pi$ interaction.
The chiral extrapolation of $f_B$ and $f_{B_s}$ including
the chiral logarithm is shown in
Figure~\ref{fig:fsqrtm_strange} with two representative 
values of $g$.
As in the pion decay constant, the uncertainty in the chiral
limit is enhanced by the chiral logarithm.

The ratio $f_{B_s}/f_B$ is needed in the extraction of the
CKM matrix element $|V_{td}/V_{ts}|$. 
Since the bulk of systematic errors cancels in the ratio,
one may expect better accuracy than the determination of
$|V_{td}|$ solely from $\Delta M_d$.
Our preliminary result varies from 1.24 (quadratic fit) to
1.38 (chiral log, $g$ = 0.59). 
It suggests that the ratio and its error can be significanly
larger than the previous world average 1.16(4).
Similar discussion, but using quenched data, has been
made in \cite{Kronfeld:2002ab}.

\section{Conclusions}
The chiral logarithm expected from ChPT is not
observed in the unquenched lattice data with $m_{PS}$ greater
than about 500~MeV, which suggests that ChPT may only be
applied in smaller mass regions.
The estimate using model functions for the chiral
extrapolation leads to the uncertainty as large as $\pm$10\%
for the decay constants.

\vspace{6mm}
\noindent
This work is supported by the KEK Supercomputer project
No.~79 (FY 2002), and also in part by the Grant-in-Aid of 
the Ministry of Education 
(Nos. 11640294, 12640253, 12740133, 13135204, 13640259, 13640260,
14046202, 14740173).
N.Y. is supported by the JSPS Research Fellowship.

\end{document}